Title: The Gauge Technique in QED_{2+1}
Author: Y.Hoshino
Comments: 9 pages , Latex, no figures , Abstract is modified

\documentclass{article}

\begin{document}

\author{Yuichi Hoshino \\
Kushiro National College of Technology \\
Otanoshike nishi-2-32-1, Kushiro, 084 Hokkaido, Japan}
\title{{\LARGE The Gauge Technique in QED}$_{2+1}$}
\maketitle

\begin{abstract}
The Gauge Technique has been applied to QED$_{2+1}$ in the quenched case
with infrared subtraction. The behaviour of the fermion propagator near the
threshold is then found to be 
\[
S(p)\rightarrow \frac{(\gamma \cdot p+m)}{(p^{2}-m^{2})}(\frac{p^{2}-m^{2}}{%
2m^{2}})^{\zeta }\exp (-\frac{\eta \varsigma }{2}), 
\]
where $\varsigma =e^{2}/(4\pi m)$ and this is gauge invariant except the
exponential factor. We also find a spectral function in the Landau and
Yennie like gauge. The propagators $S(p)$ are expressed in terms of $\Phi
(z,1,\varsigma )$ explicitly .The vacuum expectation value $\left\langle 
\overline{\psi }\psi \right\rangle $ is gauge independent but divergent .
Thus dynamical mass generation does not occur.
\end{abstract}

\newpage

\section{Introduction}

The non-linear integral Dyson-Schwinger equation has been extensively
analysed with a particular vertex ansatz or in the quenched approximation to
deal with the dynamical symmetry breaking in Quantum Field Theory. Sometimes
the approximations made do not satisfy the Ward-Takahashi (W-T) Identity. In
any case one may use the divergence of the axial-vector current to show
dynamical chiral symmetry breaking via the axial Ward identity, 
\[
\partial _{\mu x}\langle T(J_{5}^{\mu }(x)\overline{\psi }(y)\gamma _{5}\psi
(y))\rangle =-2\delta (x-y)\langle \overline{\psi }\psi \rangle \,\,\mathrm{%
\ \ and}\,\langle \overline{\psi }\psi \rangle =-\mathrm{tr}(S_{F}(x)) 
\]
in QED$_{4,3}$ and QCD$_{4}$, where the right hand side of the above
equation depends on the dynamics. Thus an effective mass, induced by gauge
interaction of the fermions, with a non vanishing $order$ parameter $\langle 
\overline{\psi }\psi \rangle \neq 0$ has been found which is similar to the
gap equation in superconductivity [4]. This is the familiar scenario of
dynamical symmetry breaking. But there remain ambiguities as they consider
only continuum contributions in Euclidian space of the fermion self energy
and the structure of the propagator is not clear. Thus we have an interest
to see what type of solutions exist in Minkowski space which satisfy the
gauge identities. We shall discuss the structure of the fermion propagator
in Minkowski space using the gauge technique, which obeys the vertex W-T
identity. The Gauge Technique, which is based on dispersion relations, leads
to a linear Dyson-Schwinger equation that admits an analytic solution [3].
On the other hand in the ladder approximation and in the Landau gauge, W-T
is valid only at one loop level of perturbation theory in non-linear
Dyson-Schwinger equation apprroach. Atkinson and Blatt[2] have studied the
singurality of the propagator after an analytic continuation of the
Dyson-Schwinger equation to Minkowski space; the only physically meaningful
answer for the propagator is a branch point(cut) on the real axis associated
with zero mass photon, but the analytic continuation from Euclidean to
Minkowski space is not unique and the result depends on the way the vertex
is treated; sometimes it leads to a complex singurality [2]. Hereafter we
confine ourselves in QED$_{2+1}.$ Because of infrared divergences in QED$%
_{2+1}$, the infrared behaviour was modified [1,2] by introducing massless
fermions into the photon vaccum polarization since it affects the low energy
behaviour of the photon. In this paper, we analyse a linearized form of the
Dyson-Schwinger equation for the fermion propagator in quenched
approximation using the gauge technique, where we treat the 2-spinor
representation of fermions in (2+1) dimensions and we do not introduce
massless sources [1,2]. In our analysis, the structure of the propagator has
an essential singurality at $p^{2}=m^{2}$ in arbitrary gauges. We also
examine the problem of dynamical mass generation and find that the vacuum
expectation value $\left\langle \overline{\psi }\psi \right\rangle $ is
gauge independent but divergent . We conclude that dynamical mass generation
does not occur .

\section{Zeroth Gauge approximation}

W-T identities between Green function in gauge theory are well known. Thus,
with photon legs amputated, the first few identities read 
\[
k^{\mu }S(p)\Gamma _{\mu }(p,p-k)S(p-k)=S(p-k)-S(p), 
\]
\begin{equation}
k^{\mu }S(p^{\prime })\Gamma _{\nu \mu }(p^{\prime }k^{\prime};pk)S(p)
\!=\!S(p^{\prime })\Gamma _{\nu }(p^{\prime },p^{\prime }\!+\!k^{\prime})
S(p^{\prime }\!+\!k^{\prime })\!-\!S(p\!-\!k^{\prime})\Gamma _{\nu }
(p\!-\!k^{\prime},p)S(p)
\end{equation}
where $S$ is a complete electron propagator and $\Gamma $ stands for the
fully amputated connected Green function; coupling constants have been
factorized out of eq(1) . In QED the propagators $S$ and $D$ of fermion and
photon, and the vertex part $\Gamma _{\mu }$ play a central role via
Dyson-Schwinger equations 
\begin{equation}
1=Z_{2}(\gamma \cdot p-m+\delta m)S(p)-ie^{2}Z_{2}\int \frac{d^{n}k}{(2\pi
)^{n}}S(p)\Gamma _{\mu }(p,p-k)S(p-k)\gamma _{\nu }D^{\mu \nu }(k)
\end{equation}
\begin{equation}
D_{\mu \nu }^{-1}(k)=Z_{3}[k^{2}g_{\mu \nu }\!-\!k_{\mu }k_{\nu }(1\!-\!\eta
^{-1})]+ie^{2}Z_{2}Tr\int\!\frac{d^{n}p}{(2\pi )^{n}}\gamma _{\nu
}S(p)\Gamma _{\mu }(p,p-k)S(p-k)
\end{equation}
\begin{equation}
\Gamma _{\mu }(p,p-k)=Z_{2}\gamma _{\mu }-ie^{2}Z_{2}\int \frac{
d^{n}p^{\prime }}{(2\pi )^{n}}\gamma _{\lambda }S(p^{\prime })\Gamma _{\nu
\mu }(p^{\prime }k^{\prime };pk)D^{\lambda \nu }(k^{\prime }).
\end{equation}
where $\eta $ is a covariant gauge parameter and we are working in $n$
-dimensions at this stage.

In the gauge technique one seeks solutions to equations (2)-(4) in the form
given above. To this end, begin with the Lehmann-Kallen spectral
representation for the fermion propagator in the form 
\begin{equation}
S(p)=(\int_{-\infty }^{-m}+\int_{m}^{\infty })dw\rho (w)\frac{1}{\gamma
\cdot p-w+i\epsilon \epsilon (w)}
\end{equation}
with 
\[
\rho (w)=\epsilon (w)\rho (w)\quad \mathrm{where}\quad \epsilon (w)=\theta
(w)-\theta (-w). 
\]
Since 
\begin{equation}
S(p-k)-S(p)=\int dw\rho (w)\frac{1}{\gamma \cdot p-w}\gamma \cdot k\frac{1}{
\gamma \cdot (p-k)-w},
\end{equation}
the simplest possible (but by no means unique) solution of (1) is to take 
\begin{equation}
S(p^{\prime })\Gamma _{\mu }^{(0)}(p^{\prime },p)S(p)=\int dw\rho (w)\frac{1%
}{\gamma \cdot p^{\prime }-w}\gamma _{\mu }\frac{1}{\gamma \cdot p-w}.
\end{equation}
The above formula represents a bare vertex weighted by a spectral function $%
\rho (w)$ for an electron of mass $w$. If the ansatz for the vertex in
equation (7) is used in quenched approximation, equation (2) is written in
three dimensions as 
\begin{eqnarray}
Z_{2}^{-1} &=&(\gamma \cdot p-m_{0})S(p)-ie^{2}\int \frac{d^{3}k}{(2\pi )^{3}%
}S(p)\Gamma _{\mu }^{(0)}(p,p-k)S(p-k)D^{\mu \nu }(k)  \nonumber \\
&=&(\gamma \cdot p-m_{0})\int \frac{\rho (w)dw}{\gamma \cdot p-w}-ie^{2}\int 
\frac{d^{3}k}{(2\pi )^{3}}dw\rho (w)\frac{1}{\gamma \cdot p-w}\gamma _{\mu }
\nonumber \\
&&\times \frac{1}{\gamma \cdot (p-k)-w}\gamma _{\nu }\cdot (g_{\mu \nu }-%
\frac{k_{\mu }k_{\nu }}{k^{2}}(1-\eta ))\frac{1}{k^{2}}  \nonumber \\
&=&\int \frac{\rho (w)dw}{\gamma \cdot p-w}(\gamma \cdot p-m_{0}+\Sigma
(p,w)),
\end{eqnarray}
where $\Sigma (p,w)$ is obtained from lowest-order self energy of the
fermion with mass $w$. Recalling $Z_{2}^{-1}=\int \rho (w)dw,$ equation (8)
can be written in the renormalized form 
\begin{equation}
0=\int \rho (w)dw\frac{w-m_{0}+\Sigma (p,w)}{\gamma \cdot p-w+i\epsilon (w)}
=\int \rho (w)dw\frac{w-m+\Sigma (p,w)-\Sigma (w,w)}{\gamma \cdot
p-w+i\epsilon (w)}.
\end{equation}
Taking the imaginary part of equation (9) yields the integral equation for
the spectral function if we replace $\gamma \cdot p=w$ $,$%
\begin{equation}
\epsilon (w)(w-m)\rho (w)=\frac{1}{\pi }\int \rho (w^{\prime })dw^{\prime }\,%
\frac{\Im \sum (w,w^{\prime })}{w-w^{\prime }}.
\end{equation}

QED$_{2+1}$ is a super renormalizable theory but there are infrared
singularities. The threshold cut opens at the $p=m_{0}$. We may carry out a
subtraction at some point, rewriting $(w-m_{0}+\Sigma (p,w))\rightarrow
(w-m+\sum (p,w)-\Sigma (w,w))$, where $m_{0}=m+\Sigma (w,w)$ represents a
mass renormalization within the integral.

\section{Zeroth Green function}

Now the self energy of the fermion can be written as $\Sigma (p,m)=\gamma
\cdot p\Sigma _{1}(p,m)+m\Sigma _{2}(p,m)$. ($\Sigma _{1}$ and $\Sigma _{2}$
are often referred to as the vector and scalar parts of the self-energy.) In
the dispersion integral for the fermion self-energy

\[
\Sigma (p,m)=\frac{1}{\pi }\int_{-\infty }^{\infty }\frac{dw}{\gamma \cdot
p-w+i\epsilon }\Im \Sigma (w,m), 
\]
$\Sigma _{1}$ is given from the even function of $w$ and $\Sigma _{2}$ is
given from odd function of $w$ in $\Im \Sigma (w,m)$. It is conventional to
replace $\gamma \cdot p=w$ in $\Im \Sigma (p,m)$. In (2+1) dimensions to $%
O(e^{2})$ and in the photon gauge specified by $\eta $ , the self energy of
the fermion is written as follows, 
\[
\Im \Sigma (w,m)=\frac{e^{2}}{16}\theta (w^{2}-m^{2})[\frac{\eta
w(w^{2}+m^{2})}{\sqrt{w^{2}}w^{2}}-\frac{2(2+\eta )m}{\sqrt{w^{2}}}]\qquad 
\mathrm{so} 
\]
\[
\Im \Sigma _{1}(w,m)=\frac{e^{2}}{16}\theta (w^{2}-m^{2})\frac{\eta
(w^{2}+m^{2})}{\sqrt{w^{2}}w^{2}}, 
\]
\begin{equation}
\Im \Sigma _{2}(w,m)=-\frac{e^{2}}{16}\theta (w^{2}-m^{2})\frac{2(2+\eta )}{%
\sqrt{w^{2}}}.
\end{equation}

If we substitute the above expression into the equation (10) it can be
expressed to this order, 
\begin{equation}
\epsilon (w)(w-m)\rho (w)=\frac{e^{2}}{16\pi }\int \frac{(\eta
w(w^{2}+w^{\prime 2})-2(2+\eta )w^{2}w^{\prime })\rho (w^{\prime
})dw^{\prime }}{\sqrt{w^{2}}w^{2}(w-w^{\prime })}.
\end{equation}
We find that the integral diverges at the $w=w^{\prime }$; however $instead$
of introducing massless source to modify the photon propagator we can remedy
the infrared divergence by replacing the selfenergy $\sum (p,m)\rightarrow
\sum (p,m)-\sum (m,m);$. Thus we subtract :$\Im \Sigma (w,w^{\prime
})-\epsilon (w)\Im \sum (w^{\prime },w^{\prime })$ to avoid the infrared
divergence (i.e.removing points at $\sqrt{p^{2}}=m)$. The integrand is
modified in the following way 
\[
\frac{1}{w-w^{\prime }}\eta ((1+\frac{w^{\prime 2}}{w^{2}})-2)-2(2+\eta )(%
\frac{w^{\prime }}{w}-1)=\frac{(4+\eta )}{w}-\eta \frac{w^{\prime }}{w^{2}}. 
\]
In this way the spectral function obeys the more sensible integral equation 
\begin{eqnarray}
\epsilon (w)(w\!-\!m)\rho (w) &=&\frac{\xi (\eta +4)}{w}(\int_{m}^{w\epsilon
(w)}\!-\!\int_{-w\epsilon (w)}^{-m})\rho (w^{\prime })dw^{\prime }  \nonumber
\\
&&-\frac{\xi \eta }{w^{2}}(\int_{m}^{w\epsilon (w)}\!-\!\int_{-w\epsilon
(w)}^{-m})w^{\prime }\rho (w^{\prime })dw^{\prime },
\end{eqnarray}
where $\xi =\frac{e^{2}}{16\pi }$. (The 3-D analogue of the 4-D Yennie type
gauge is $\eta =-4$.) The infrared behaviour of the solution is given by
multiplying $w^{2}$ and set $w=m$ in front of the integral in the right hand
side of equation (13) thereby converting it into the approximate
differential equation 
\begin{eqnarray*}
m^{2}\frac{d}{dw}[(w-m)\rho (w)] &\simeq &m\xi (\eta +4)(\rho (w)-\rho
(-w))-\xi \eta w(\rho (w)+\rho (-w)) \\
m^{2}\frac{d}{dw}[(w+m)\rho (-w)] &\approx &m\xi (\eta +4)(\rho (-w)-\rho
(w))-\xi \eta w(\rho (w)+\rho (-w)),
\end{eqnarray*}
from which we obtain near $w=m$%
\begin{equation}
\rho (\frac{w}{m})-\rho (-\frac{w}{m})\approx \exp (-\frac{\varsigma }{2}
\eta \frac{w}{m})(\frac{w}{m}-1)^{-1+\varsigma }(\frac{w}{m}
+1)^{-1-\varsigma },
\end{equation}
then 
\begin{equation}
\rho (\frac{p}{m})\equiv \rho (\frac{p}{m})-\rho (-\frac{p}{m})\approx (%
\frac{p^{2}-m^{2}}{2m^{2}})^{-1+\varsigma }\exp (-\frac{\varsigma \eta }{2}
),\qquad
\end{equation}
\begin{equation}
S(p)_{p^{2}\simeq m^{2}}\rightarrow \frac{(\gamma \cdot p+m)}{(p^{2}-m^{2})}(%
\frac{m^{2}}{p^{2}-m^{2}})^{-\zeta }\exp (-\frac{\varsigma \eta }{2}),\quad
\zeta =\frac{e^{2}}{4\pi m}.
\end{equation}
This is a new feature of QED$_{2+1}$. It shows the gauge independence near
the threshold. There is no special gauge in which the fermion has a free
pole near $p^{2}=m^{2}$ . In QED$_{3+1}$, the Yennie gauge $\eta =-3$
produces a free particle pole (Abrikosov 1956) in the lowest approximation
[3] : 
\begin{equation}
S(p)_{p^{2}\simeq m^{2}}\sim \frac{(\gamma \cdot p+m)}{(p^{2}-m^{2})}\left( 
\frac{m^{2}}{p^{2}-m^{2}}\right) ^{\alpha (\eta -3)/2\pi }.
\end{equation}

It is often helpful to split the spectral function in odd and even parts by
[1,3] 
\[
\rho (w)=\epsilon (w)[w\rho _{1}(w)+m\rho _{2}(w)]. 
\]
But in our case this is unnecessary, as shown below. Introducing
dimensionless variables, $\frac{w}{m}=\omega ,\varsigma =\frac{e^{2}}{4\pi m}%
,$ we rewrite the equation (13) as 
\begin{eqnarray}
\epsilon (\omega )(\omega \!-\!1)\rho (\omega ) &=&\frac{\varsigma (\eta +4)%
}{\omega }(\int_{1}^{\omega \epsilon (\omega )}\!-\!\int_{-\omega \epsilon
(\omega )}^{-1})\rho (\omega ^{\prime })d\omega ^{\prime }  \nonumber \\
&&-\frac{\varsigma \eta }{\omega ^{2}}(\int_{1}^{\omega \epsilon (\epsilon
)}-\int_{-\omega \epsilon (\omega )}^{-1})\omega ^{\prime }\rho (\omega
^{\prime })d\omega ^{\prime }.
\end{eqnarray}
It is easy to see that the eqn (18) can be converted into a first order
differential equation in the Landau gauge $\eta =0$, as well as in the
Yennie like gauge $\eta =-4$. We separate the equation in the different
region of $\omega $ by $\epsilon (\omega )=\theta (\omega )-\theta (-\omega
) $ . In fact for those gauges the differential equations read 
\begin{eqnarray}
\frac{d}{d\omega }(\omega (\omega -1)\rho (\omega )) &=&\varsigma (\rho
(\omega )-\rho (-\omega )),\omega >0  \nonumber \\
\frac{d}{d\omega }(\omega (\omega +1)\rho (-\omega )) &=&\varsigma (\rho
(-\omega )-\rho (\omega )),\omega <0(\omega \rightarrow -\omega ),\eta =0,
\end{eqnarray}
\begin{eqnarray}
\frac{d}{d\omega }(\omega ^{2}(\omega -1)\rho (\omega )) &=&\varsigma \omega
(\rho (\omega )+\rho (-\omega )),\omega >0  \nonumber \\
\frac{d}{d\omega }(\omega ^{2}(\omega +1)\rho (-\omega )) &=&\varsigma
\omega (\rho (-\omega )+\rho (\omega )),\omega <0,\eta =-4
\end{eqnarray}
and the spectral function solutions are given by 
\begin{eqnarray}
\rho (\omega ) &=&-\frac{C_{1}}{\varsigma \omega (\omega -1)}-\frac{C_{2}}{
\omega (\omega -1)}(\frac{\omega -1}{\omega +1})^{\varsigma }-\frac{C_{1}}{
\omega (\omega -1)}\Phi (\frac{1+\omega }{1-\omega },1,-\varsigma ) 
\nonumber \\
\rho (-\omega ) &=&\frac{C_{1}}{\omega (\omega +1)}\Phi (\frac{1+\omega }{%
1-\omega },1,-\varsigma )+\frac{C_{2}}{\omega (\omega +1)}(\frac{\omega -1}{
\omega +1})^{\varsigma },\eta =0
\end{eqnarray}
and 
\begin{eqnarray}
\rho (\omega ) &=&\frac{C_{1}}{\omega ^{2}(\omega -1)}(\frac{\omega -1}{
\omega +1})^{\varsigma }-\frac{C_{2}}{\varsigma \omega ^{2}(\omega -1)}+%
\frac{C_{2}}{\omega ^{2}(\omega -1)}\Phi (\frac{1+\omega }{1-\omega }
,1,-\varsigma )  \nonumber \\
\rho (-\omega ) &=&\frac{C_{1}}{\omega ^{2}(\omega +1)}(\frac{\omega -1}{%
\omega +1})^{\varsigma }-\frac{C_{2}}{\varsigma \omega ^{2}(\omega +1)}+%
\frac{C_{2}}{\omega ^{2}(\omega +1)}\Phi (\frac{1+\omega }{1-\omega }%
,1,-\varsigma ),  \nonumber \\
\eta &=&-4,
\end{eqnarray}
where

\[
\int \frac{dw}{w(w-1)}\mathbf{(}\frac{w+1}{w-1}\mathbf{)}^{\varsigma }%
\mathbf{=(}\frac{w+1}{w-1}\mathbf{)}^{\varsigma }\mathbf{\Phi (}\frac{1+w}{%
1-w}\mathbf{,1,-\varsigma ),} 
\]
\[
\int_{0}^{1}\frac{t^{v-1}dt}{1-zt}=\Phi (z,1,v), 
\]
\begin{eqnarray}
\Phi (z,a,\nu ) &=&\sum_{n=0}^{\infty }\frac{z^{n}}{(\nu +n)^{a}}  \nonumber
\\
&=&\frac{1}{\Gamma (a)}\int_{0}^{\infty }\frac{t^{a-1}e^{-vt}}{1-ze^{-t}}dt.
\end{eqnarray}
In order to agree with the free field limit $\rho (\omega )\rightarrow
\delta (\omega -1)$, where the condition $\int \rho (w)dw=1$ holds, the
spectral function $\rho (\omega )$ is normalized in the Landau gauge to 
\begin{eqnarray}
\rho (\omega ) &=&\frac{\varsigma }{m}\theta (\omega -1)\frac{1}{\omega
(\omega -1)}(\frac{\omega -1}{\omega +1})^{\varsigma }  \nonumber \\
\rho (-\omega ) &=&\frac{\varsigma }{m}\theta (\omega -1)\frac{1}{\omega
(\omega +1)}(\frac{\omega -1}{\omega +1})^{\varsigma }
\end{eqnarray}
This leads to quantities $Z_{2},$ $m_{0}$ for small $\varsigma $ $,$ 
\[
Z_{2}^{-1}=\int \rho (w)dw=\int_{1}^{\infty }(\rho (\omega )-\rho (-\omega
))d\omega =1 
\]
\begin{equation}
m_{0}Z_{2}^{-1}=m\int_{1}^{\infty }\omega (\rho (\omega )+\rho (-\omega
))d\omega =m.
\end{equation}
More generally

\[
\int \rho (\omega )f(\omega )d\omega =\int_{1}^{\infty }(\rho (\omega
)f(\omega )-\rho (-\omega )f(-\omega ))d\omega . 
\]
Therefore 
\begin{equation}
S(p)=2\varsigma \int_{1}^{\infty }d\omega \frac{\gamma \cdot p+m}{
p^{2}-m^{2}\omega ^{2}}\frac{1}{\omega ^{2}-1}(\frac{\omega -1}{\omega +1}
)^{\varsigma },
\end{equation}
it can be expressed in terms of the higher transcendental function 
\[
\Phi (\frac{p-m}{p+m},1,\varsigma )\quad {,}\Phi (\frac{p+m}{p-m}
,1,\varsigma ), 
\]
and 
\begin{equation}
S(p)=\frac{\gamma \cdot p+m}{p^{2}-m^{2}}(1+\varsigma \epsilon (p)\Phi (%
\frac{p+m}{p-m},1,\varsigma )).
\end{equation}
Notice that the point $p=\infty $ corresponds to a branch point at $z=1$ and
the point $p=m$ to $z=\infty $ or $0$ . In the gauge $\eta =-4$, the
spectral function with correct normalization is instead given by 
\begin{eqnarray}
\rho (\omega ) &=&\frac{\varsigma }{m}\theta (\omega -1)\frac{1}{\omega
^{2}(\omega -1)}(\frac{\omega -1}{\omega +1})^{\varsigma }  \nonumber \\
\rho (-\omega ) &=&\frac{\varsigma }{m}\theta (\omega -1)\frac{1}{\omega
^{2}(\omega +1)}(\frac{\omega -1}{\omega +1})^{\varsigma }.
\end{eqnarray}
leading to 
\[
Z_{2}^{-1}=1+2\varsigma -4\varsigma ^{2}\Phi (-1,1,\varsigma )\rightarrow 1 
\]
\begin{equation}
m_{0}Z_{2}^{-1}=m
\end{equation}
and 
\begin{equation}
S(p)=2\varsigma \int_{1}^{\infty }d\omega \frac{1}{p^{2}-m^{2}\omega ^{2}}(%
\frac{\gamma \cdot p}{\omega ^{2}(\omega ^{2}-1)}+\frac{m}{\omega ^{2}-1})(%
\frac{\omega -1}{\omega +1})^{\varsigma }
\end{equation}

Therefore it is written as

\begin{eqnarray}
S(p) &=&\frac{\gamma \cdot p}{p^{2}-m^{2}}[1+\frac{\varsigma m^{4}}{p^{3}}%
\epsilon (p)(\frac{p+m}{p-m}\Phi (\frac{p-m}{p+m},1,\varsigma ))]  \nonumber
\\
&&-2\varsigma \frac{\gamma \cdot p}{p^{2}}(1-2(\varsigma -1)\Phi
(-1,1,\varsigma ))  \nonumber \\
&&+\frac{m}{p^{2}-m^{2}}[1+\varsigma \epsilon (p)\Phi (\frac{p-m}{p+m}%
,1,\varsigma )
\end{eqnarray}
It is interesting to examine the possible occurrence of confinement and
dynamical mass generation in our model. In general $Z_{2}=0$ is a
compositeness condition and $m_{0}Z_{2}^{-1}=0$ is a signpost of dynamical
mass generation. However in our case two conditions are not satisfied . But
the vacuum expectation value $\left\langle \overline{\psi }\psi
\right\rangle $ is a gauge invariant quantity in general. We examine the
order parameter in the above two gages and find

\begin{eqnarray}
\left\langle \overline{\psi }\psi \right\rangle &=&-itrS(x)=2\int \frac{
d^{3}p}{(2\pi )^{3}}\int dw\rho (w)\frac{w}{p^{2}+w^{2}}  \nonumber \\
&=&-\infty (\eta =0,-4)
\end{eqnarray}

Here we used the dimensional reguralization . In our case the order
parameter is gauge independent but divergent. We deduce that in our
approximation there is no dynamical mass generation .

\section{Improved Vertex function}

Direct substitution of our solution $\rho (w)$ in two gauges (21) and (22)
into equation $(7)$, gives the zeroth the vertex function, 
\begin{equation}
S(p)\Gamma _{\mu }^{(0)}S(p^{\prime })=\int dw\rho (w)\frac{1}{\gamma \cdot
p-w}\gamma _{\mu }\frac{1}{\gamma \cdot p^{^{\prime }}-w}.
\end{equation}
We may improve the vertex by adding transverse combinations of terms such as 
\[
\lbrack (p_{\mu }\gamma \cdot p^{^{\prime }}+p_{\mu }^{^{\prime }}\gamma
\cdot p-p\cdot p^{^{\prime }}\gamma _{\mu })+((p+p^{^{\prime }})_{\mu
}+i(p^{^{\prime }}-p)_{\alpha }\epsilon _{\alpha \mu \nu }\gamma _{\nu
})w+w^{2}\gamma _{\mu }] 
\]
in the numerator of the integral, allowing for a parity violating
contribution. One may also include a form factor proportional to a linear
combination of $\gamma _{\mu }$ and $(p+p^{^{\prime }})_{\mu }$ and try to
ensure that on-shell quantities are gauge invariant.

\section{Summary}

Previously, to soften the infrared divergence, massless fermions were
introduced into photon vacuum polarization [1]. In this work we have instead
analyzed the quenched case and made an infrared subtraction to avoid the
infrared singurality. We obtained the solution of the spectral function $%
\rho (\omega )$ and the momentum space propagator $S(p)$ in the Landau gauge
and in the Yennie like gauge ($\eta =-4)$; these are simple functions like
in $QED_{3+1}[3].$ Here we summarize the differences between two
approximations. The former shows the gauge invariance of the spectral
function near the mass shell, the high energy behaviour of the spectral
function is $1/p^{2}$ and the cut structure near the mass shell is given by
massless fermion-loop ; $\rho (p)$ $=(p-m)^{-1-e^{2}/c\pi ^{2}}$ for $p-m\ll
e^{2}$ and $c=e^{2}N/8$ for $N$ massless fermions; the integral equation for 
$\rho (\omega )$ was not solved analytically because of its complexity, due
to higher order corrections. In our approximation the high energy behaviour
is $1/p^{2}$ in the Landau gauge and $1/p^{4}$ in the Yennie like gauge, and
structure near the mass shell is determined by coupling constant mass ratio $%
\varsigma =e^{2}/(4\pi m)$. Most importantly we arrived at is the gauge
independence of the cut near the mass shell $p=m$ . The order parameter is
gauge independent but divergent , so there is no dynamical mass generation .
In contrast to (3+1) dimensions, the gauge dependence near the mass shell is
very weak except for the exponential factor . Since $QED_{2+1}$ is super
renormalizable, ordinary renormalization group is not relevant, but the
gauge technique is non-perturbative, and we have succeeded in finding the
infrared behaviour of the fermion propagator as well as its full structure
in terms of $\Phi (z,1,\varsigma )$ .

\section{Acknowledgement}

The author is indebted to Prof. Delbourgo during his stay in March 2000 at
the University of Tasmania for his hospitality and stimulating comments. He
also thanks to Dr.Suzuki for his careful comment in the differential
equation .

\section{References}

\noindent [1] T. Appelquist, D. Nash and L.C.R. Wijewardhana, Phys. Rev.
Lett. \textbf{60} (1988) 2575; A.B. Waites and R. Delbourgo, Int. J. Mod.
Phys. \textbf{27A} (1992) 6857; Y. Hoshino and T. Matsuyama, Phys. Lett. 
\textbf{222B} (1989) 493.

\noindent [2] D. Atkinson D. and D.W.E. Blatt, Nucl. Phys. \textbf{151B}
(1979) 342; P. Maris, Phys. Rev. \textbf{52D} (1995) 6087; Y. Hoshino, Nuovo
Cim. \textbf{112 A} (1999) 335.

\noindent [3] Abrikosov, A.A,JETP,\textbf{\ 30},96(1956); R. Delbourgo and
P. West, J. Phys. \textbf{10A} (1977) 1049, Phys. Lett. \textbf{B72B} (1977)
96; R. Delbourgo, Nuovo Cim. \textbf{49A} (1979) 484.

\noindent [4] T. Maskawa and H. Nakajima, Prog. Theor. Phys. \textbf{52}
(1974) 1326, Prog. Theor. Phys. \textbf{54} (1975) 860; R. Fukuda and T.
Kugo, Nucl. Phys. \textbf{B117} (1976) 250; H.D. Politzer, Nucl. Phys. 
\textbf{B117} (1976) 397.

\noindent [5] S. Deser, R. Jackiw and S. Templeton, Ann . Phys. \textbf{140 }
(1982) 372.

\noindent

\end{document}